\documentclass[%
 aip,
 amsmath,amssymb,
 reprint,%
]{revtex4-1}

\usepackage{graphicx}
\usepackage{dcolumn}
\usepackage{bm}

\usepackage[utf8]{inputenc}
\usepackage[T1]{fontenc}
\usepackage{mathptmx}
\usepackage{etoolbox}

\makeatletter
\def\@email#1#2{%
 \endgroup
 \patchcmd{\titleblock@produce}
  {\frontmatter@RRAPformat}
  {\frontmatter@RRAPformat{\produce@RRAP{*#1\href{mailto:#2}{#2}}}\frontmatter@RRAPformat}
  {}{}
}%
\makeatother
\begin{document}

\preprint{AIP/123-QED}

\title[Sample title]{2.34 kV $\mathrm{\beta} $-Ga\textsubscript{2}O\textsubscript{3}  Vertical Trench RESURF Schottky Barrier Diode with sub-micron fin width}
\author{Chinmoy Nath Saha}
\author {Saurav Roy}%
\author {Yizheng Liu}
\author {Carl Peterson}
\author {Sriram Krishnamoorthy}%
%

\affiliation{%
Materials Department, University of California Santa Barbara, Santa Barbara-93106
}%

\date{\today}

\begin{abstract}
In this letter, we present a kilovolt-class $\mathrm{\beta} $-Ga\textsubscript{2}O\textsubscript{3} vertical trench Schottky barrier diode with a field plate incorporating narrow fin width (W\textsubscript{fin}) structures of sub-micron dimensions. We used a nanolaminate dielectric comprising a stack of multiple thin TiO\textsubscript{2} and Al\textsubscript{2}O\textsubscript{3} layers as RESURF dielectric and for field plate edge termination. Both W\textsubscript{fin} of 200 nm and 500 nm demonstrate excellent on-state performance with specific on-resistance (R\textsubscript{on,sp}) of 9.8–12 m$\mathrm{\Omega}$cm\textsuperscript{2},  and 10\textsuperscript{10} rectification ratio. A self-aligned photoresist planarization and etch-back process was employed to expose the top of the fins for Schottky contact formation, eliminating critical lithographic alignment challenges in sub-micron scale processing. We achieved a breakdown of 2.34 kV with very low leakage currents before catastrophic breakdown. The measured breakdown voltage is limited by dielectric breakdown at the trench bottom corner as verified by metal-oxide-semiconductor (MOS) test structure. 
TCAD simulation shows a reduced electric field at the surface of the metal-semiconductor junction due to the RESURF effect, resulting in very low reverse leakage before breakdown. The parallel plane electric field in the $\mathrm{\beta} $-Ga\textsubscript{2}O\textsubscript{3} is extracted to be 3.8 MV/cm from TCAD simulations using accurately extracted drift layer doping profile from high voltage CV measurements. A power figure of merit of 0.867 GW/cm\textsuperscript{2}(0.56 GW/cm\textsuperscript{2} with current spreading) was calculated. Enhanced RESURF by integration of high-k dielectrics with self-aligned photoresist planarization, offers a promising pathway towards high figure of merit, low leakage high-performance vertical devices.
\end{abstract}

\maketitle
$\mathrm{\beta} $-Ga\textsubscript{2}O\textsubscript{3} has attracted extensive interest in the field of ultra-wide band gap semiconductor due to its favorable material properties such as predicted high breakdown strength (8 MV/cm)\cite{green2022beta}, high saturation velocity\cite{ghosh2017ab} and electron mobility of 200 cm\textsuperscript{2}/vs \cite{peterson2024200}. Bulk crystal growth (CZ, EFZ)\cite{blevins2019development,kuramata2016high} and epitaxial growth techniques (MBE,MOCVD,HVPE)\cite{peterson2024200,sasaki2012device,sasaki2014growth,murakami2014homoepitaxial} with controllable doping have shown good prospects for low-cost wafer-scale device fabrication for high-voltage power switches and RF transistors. kV -class $\mathrm{\beta} $-Ga\textsubscript{2}O\textsubscript{3} MOSFETs\cite{kezeng2018,tetzner2019lateral,arkka1,li2019single}, Schottky\cite{farzana2023vertical,qin20232,hao2023improved}and NiO\textsubscript{x}/$\mathrm{\beta} $-Ga\textsubscript{2}O\textsubscript{3} p-n diodes\cite{li20237,wan20243,wang20212} with a breakdown field up to 5.5 MV/cm\cite{kalarickal2021beta,sahabeta2} and breakdown voltage exceeding 10 KV\cite{liu202310} have been reported in recent times. Scaled RF devices demonstrated excellent high-frequency performance \cite{zhou2025c,sahabeta2,moser2020toward,saha2025high}  using thin-channel MOSFETs and Hetero-structure FETs.

A good power rectifier requires low turn-on voltage (V\textsubscript{on}) and high current density in the on-state and the capability of blocking large voltage during the off-state. Schottky diodes have an inherent advantage of low conduction loss in the on-state because of low on-state voltage drop. But, a high electric field at the Schottky contact causes large reverse leakage through thermionic field emission and barrier-lowering effect. 

Achieving lower conduction loss by using lower work function metal comes at a compromise of exceedingly high reverse leakage currents due to reduced barrier height.  Trench Schottky barrier diodes take advantage of the RESURF (Reduced Surface Field) effect by reducing the electric field at the metal-semiconductor junction and and the high field region is buried into the drift layer. So, we can achieve lower leakage and higher breakdown field of $\mathrm{\beta} $-Ga\textsubscript{2}O\textsubscript{3} without compromising turn-on voltage and conduction loss \cite{roy2024low}. In addition, p type doping, which is yet to be realized in Ga\textsubscript{2}O\textsubscript{3}, is not necessary for trench RESURF structure to alleviate the concentration of electric field lines at device
edges and corners. 
Trench Schottky barrier diodes with breakdown voltage up to 3 kV \cite{roy2023ultra,roy2024low}  have been reported in literature with the lowest fin widths upto 1 $\mu$m. \cite{li20182,li2019field,roy2023ultra,roy2024low,sasaki2017first,dhara2023beta,dhara2025charge}. The reverse leakage current of trench Schottky diodes before breakdown has been reported to be much lower compared to planar Schottky diodes, demonstrating the benefit of using trench geometry in a power rectifier. However, the final breakdown voltage of the diode is possibly limited by the dielectric breakdown at the trench bottom corner where the highest electric field is concentrated. High-k dielectric such as BaTiO\textsubscript{3} can be an effective solution for this technique because of the very low potential drop in the dielectric. This can enable us to achieve the predicted higher breakdown field in the $\mathrm{\beta} $-Ga\textsubscript{2}O\textsubscript{3} drift layer. A smaller fin width is also preferable since peak electric field at trench corners is found to decrease with
reducing fin width \cite{li20182} and the RESURF effect is enhanced for lower W\textsubscript{fin}. Simulation study\cite{roy2024low} shows that sub-micron fin width can effectively reduce the electric field at the metal-semiconductor junction to a very minimum value due to better RESURF effect compared to fin width > 1 $\mu$m. But not all dielectrics are suitable for a sub-micron trench diode. Dielectric has to be removed from the top of the sub-micron fin with precise alignment during lithography presenting challenges for the realization of a high-yield process.

In this letter, we developed a self-aligned photoresist planarization and etch-back-based process to incorporate sub-micron fin width dimension in the trench diode fabrication without the need for any critical lithography alignment. We used a dielectric composed of thin layers of TiO\textsubscript{2}/Al\textsubscript{2}O\textsubscript{3} which has higher dielectric constant than Al\textsubscript{2}O\textsubscript{3} and also has higher etch selectivity than photoresist during the dry plasma etch process unlike BaTiO\textsubscript{3}. We achieved an excellent on-state performance with low R\textsubscript{on,sp}, decent current density of 140-165 A/cm\textsuperscript{2} and 10\textsuperscript{10} on-off ratio. Both 100 $\times$ 100 $\mathrm{\mu}$m\textsuperscript{2} and 200 $\times$ 200 $\mathrm{\mu}$m\textsuperscript{2} device show low reverse leakage before breakdown at 2.34 kV resulting in 0.867 GW/cm\textsuperscript{2} Power figure of merit (PFoM).

\begin{figure*}
\centerline{\includegraphics[width=1.8\columnwidth]{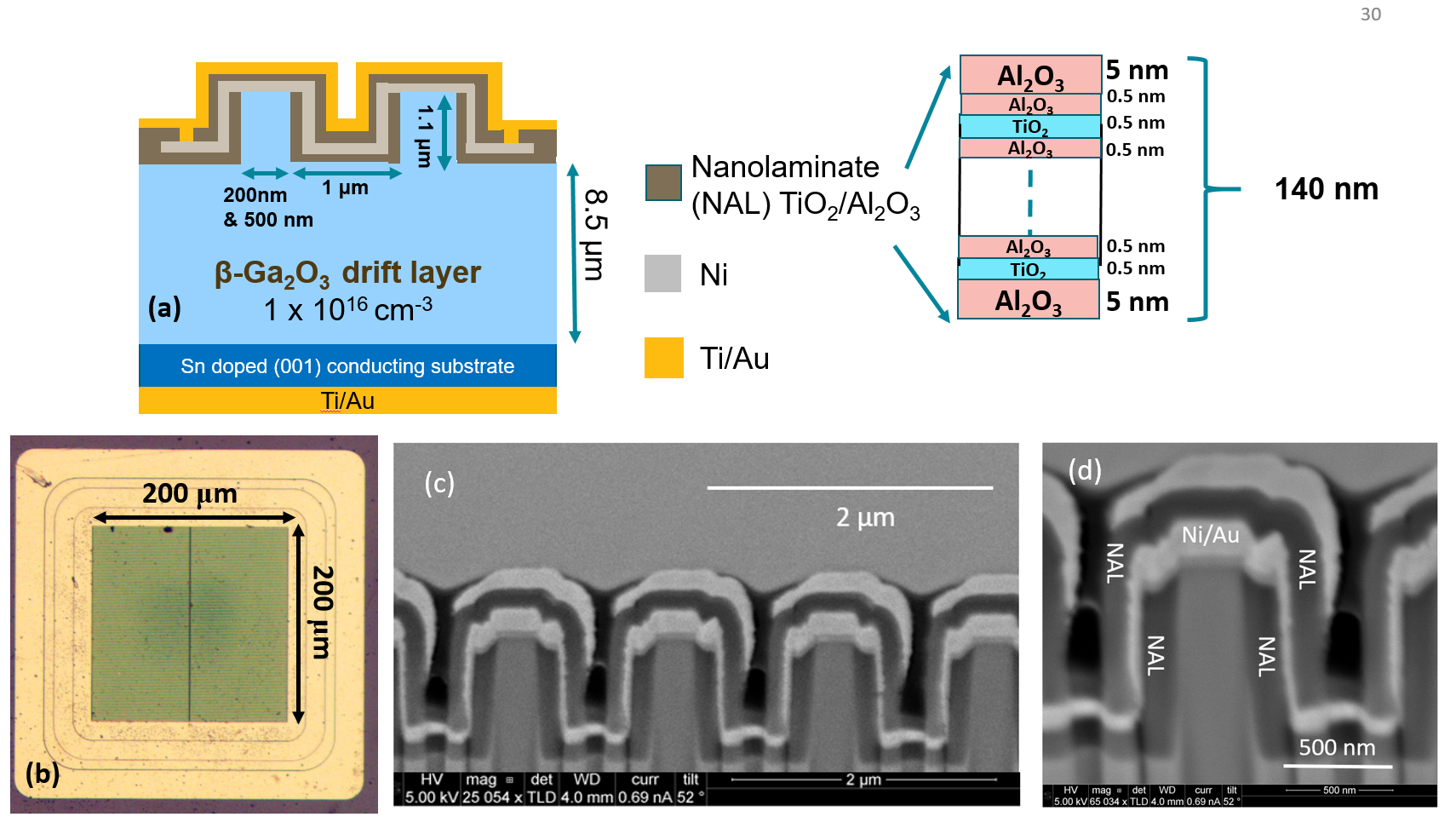}}
\caption{(a) Cross-section schematic of the final device structure of the trench Schottky barrier diode, (b) Optical microscope view of a fabricated 200 $\times$ 200 $\mu$m\textsuperscript{2} device (c) FIB-SEM image of a fabricated device with W\textsubscript{fin} = 200 nm showing multiple fin and (d) single fin (enlarged view).}
\label{fig1 crosssection}
\end{figure*}

The epitaxial structure of the device consists of 8.5 $\mu$m (001) oriented HVPE grown drift layer (expected doping 10\textsuperscript{16}cm\textsuperscript{-3}) on top of Sn-doped conducting substrate provided by Novel Crystal Technology (NCT), Japan. We defined fin dimension (W\textsubscript{fin}) of 200 nm and 500 nm using Electron Beam Lithography (EBL). The dimension of the pitch was around  1.2  $\mu$m and 1.5 $\mu$m for W\textsubscript{fin} of 200 nm and 500 nm device, respectively. Approximately 1 $\mu$m of the drift layer was dry etched by BCl\textsubscript{3} chemistry based ICP-RIE (Inductively Coupled Plasma- Reactive Ion Etching) using Ru/SiO\textsubscript{2} hard mask. The sample was submerged in the 48\% HF solution\cite{li2019field} for 10 min followed by HCL\cite{bhattacharyya2022high} for 15 min to minimize any dry etch induced damage in the sidewall and clean the interface before dielectric deposition. 
140 nm dielectric consisting of thin (0.5 nm)
 alternating layer of TiO\textsubscript{2}/Al\textsubscript{2}O\textsubscript{3} (known as nanolaminate (NAL)\cite{li2010giant,lee2013interface}) was deposited using plasma ALD. The ALD dielectric was capped by 5 nm  Al\textsubscript{2}O\textsubscript{3} on both top and bottom to  add conduction band offset at the dielectric/$\mathrm{\beta} $-Ga\textsubscript{2}O\textsubscript{3} interface. 2.8 $\mu$m thick photoresist was spin-coated and soft baked for the planarization process. We used O\textsubscript{2} plasma to etch back the photoresist until fin heads were visible on Scanning Electron Microscopy (SEM). Dielectric on top of the fins was etched using BCl\textsubscript{3} based dry etch as mentioned earlier. Due to higher selectivity, the remaining photoresist (400-600 nm) covering the nanolaminate on the trench bottom and sidewall was not etched when the nanolaminate was etched from the top of the fins. After photoresist removal, Ni/Au anode contact was deposited using planetary rotation for conformal anode metal deposition to cover top and sidewall. We deposited 110 nm nanolaminate using plasma ALD for the field-plate oxide to reduce edge field crowding. The nanolaminate was selectively etched using the similar method as described before at the edges of the device for field plating. Field plate metallization was performed using e-beam evaporation. Finally Ohmic Ti/Au contacts were deposited on the back-side of the doped substrate. The cross-section schematic of the fabricated trench diode is shown in Fig \ref{fig1 crosssection} (a), optical image  for a 200 $\times$ 200 $\mu$m\textsuperscript{2} device is shown in Fig \ref{fig1 crosssection}(b). Focused Ion Beam (FIB) cross-section image of a multi-fin device with W\textsubscript{fin} = 200 nm is shown in Fig \ref{fig1 crosssection} (c). A magnified image of a single-fin portion of the device is shown in Fig \ref{fig1 crosssection}(d). FIB-SEM image shows a slightly rounded corner at the bottom of the fin, which will reduce electric field crowding.

\begin{figure}[h]
\centerline{\includegraphics[width=1.0\columnwidth]{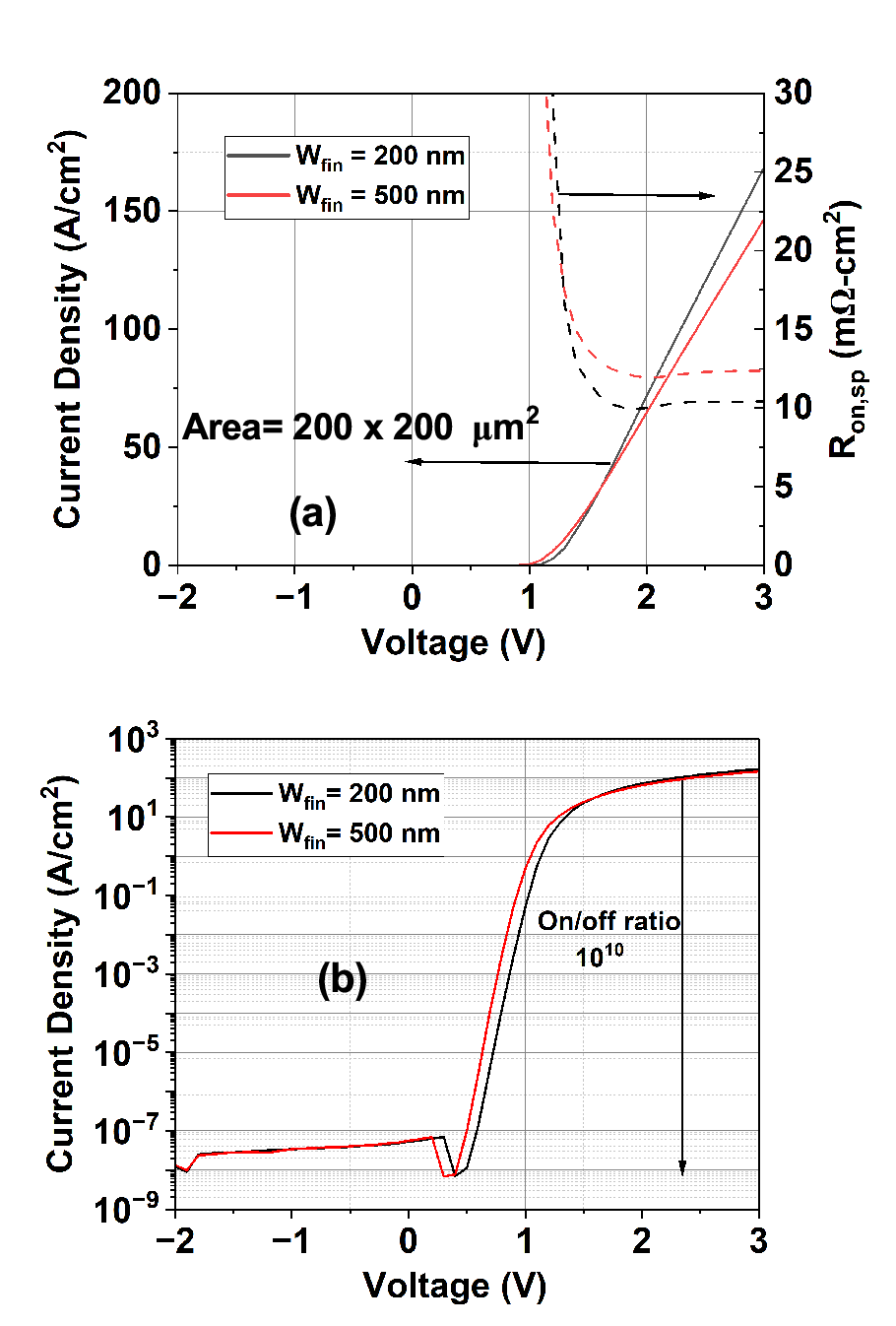}}
\vspace*{-2mm}
\caption{(a) Linear current-voltage characteristics of trench Schottky barrier diode with area = 200 $\times$ 200 $\mu$m\textsuperscript{2} and corresponding R\textsubscript{on,sp} after considering current spreading. (b) log-IV curve of trench diodes showing an on-off ratio around 10\textsuperscript{10} and ideality factor ranging from 1.07-1.18.}
\label{fig2 IV}
\end{figure}

Figure \ref{fig2 IV} (a) illustrates the linear current-voltage characteristics of trench Schottky diodes for 200 $\times$ 200 $\mu$m\textsuperscript{2} size devices. We observed a peak current density of 140-160 A/cm\textsuperscript{2} at 3 V bias for both devices. The differential specific on resistance (R\textsubscript{on,sp}) was extracted to be in the range of  9.8 - 12 m$\mathrm{\Omega}$-cm\textsuperscript{2} with current spreading taken into account. 
The specific on-resistance of the diodes after considering current
spreading is calculated by normalizing the measured absolute current
with the total spreading area (A\textsubscript{sp})\cite{yates2022demonstration}, where A\textsubscript{sp} = (Side of square +2 L\textsubscript{Drift)})\textsuperscript{2}. Here, we assume 45\textsuperscript{0} current spreading angle\cite{li2019single,roy20232} and  drift layer thickness is L\textsubscript{drift}.
Without considering current spreading, R\textsubscript{on,sp} is calculated to be 6.5- 9.89 m$\mathrm{\Omega}$-cm\textsuperscript{2} range values.

All devices independent of fin width and area have a similar on-off ratio of 10\textsuperscript{10} and  1.1 V turn-on voltage (V\textsubscript{on}). The ideality factor of all size devices with W\textsubscript{fin} = 200 nm and 500 nm range from 1.08 to 1.17, extracted from the log IV curve (Fig \ref{fig2 IV}(b)). We can define the area ratio of our trench diode by the ratio of the fin width and total pitch. The area ratio of W\textsubscript{fin} 200 nm and W\textsubscript{fin} = 500 nm device become 16\% and 33\% for respectively. Even with a lower area ratio, forward current density is not limited by sub-micron fin width and both W\textsubscript{fin} = 200 nm and 500 nm show similar current density and R\textsubscript{on,sp} (Fig \ref{fig2 IV}(a)). This can be attributed to the low power ICP RIE etch and post-etch cleaning by 48\% HF and HCl. This post-etch processing may have reduced etch damage. Besides, interface trap density could be low because of the reduced etch damage in the fin sidewall \cite{li2019field,li20182,li2020guiding}. Post Deposition annealing (PDA) inside ALD chamber for 1 hour at 250\textsuperscript{0}C has been reported to reduce defect density at the  Al\textsubscript{2}O\textsubscript{3}/$\mathrm{\beta} $-Ga\textsubscript{2}O\textsubscript{3} interface \cite{islam2022defect}. In our work, 140 nm plasma ALD dielectric deposition at 300\textsuperscript{0}C took 6 hours to complete, and the sample was kept inside the chamber for 1 additional hour. We predict that long deposition time may have unintentionally reduced the interface state density. Trapped negative charge at the sidewall and Ga\textsubscript{2}O\textsubscript{3}/dielectric interface can cause a depletion of the fin channel and increase resistance. Near-identical specific on-resistance for both sub-micron fin width dimensions indicates 
low interface state density and negligible sidewall depletion.

Li et al \cite{li2019field,li2020guiding} reported that if there is no sidewall depletion from the interface state charge density at the MOS interface, the R\textsubscript{on,sp} is approximately independent of fin width for a particular area ratio. In their predicted model, an area ratio of 33\% with similar doping and thickness of $\mathrm{\beta} $-Ga\textsubscript{2}O\textsubscript{3} drift layer compared to this report, can result in R\textsubscript{on,sp}= 8.2 m$\mathrm{\Omega}$-cm\textsuperscript{2} approximately. This is similar to the measured R\textsubscript{on,sp} in this work, which proves the sidewall depletion is negligible for both the fin widths in our work.

\begin{figure}
\centerline{\includegraphics[width=1.1\columnwidth]{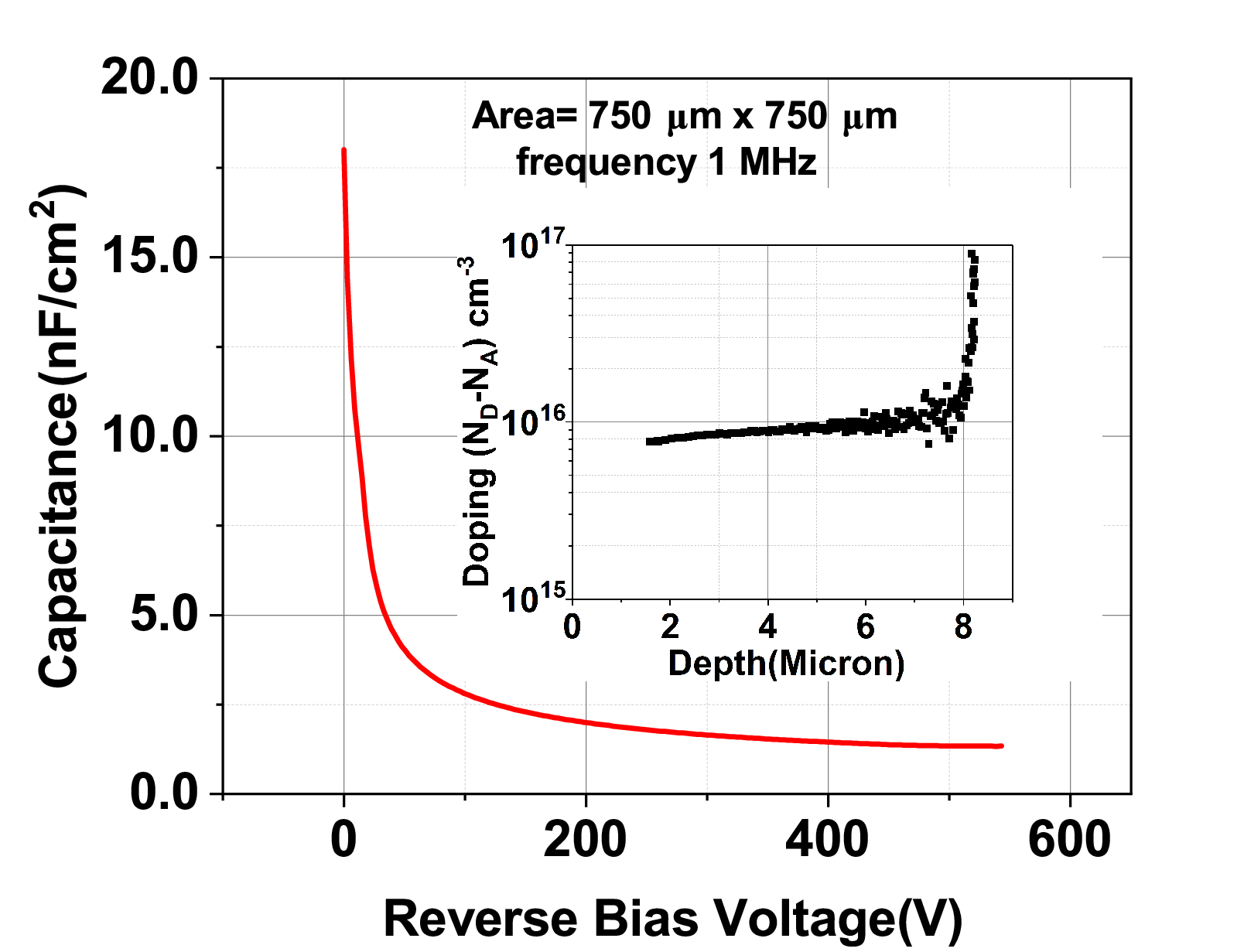}}
\vspace*{-2mm}
\caption{(a) Measured CV characteristics of 750 $\mu$m $\times$ 750 $\mu$m MOS capacitor structure showing full punch through at 550 V. Inset shows the extracted carrier density vs depth profile from CV measurement. The calculated drift layer thickness = 8 to 9 $\mu$m with average doping = 9.5 x 10\textsuperscript{15}cm\textsuperscript{-3}. }
\label{fig3 CV}
\end{figure}
\begin{figure}
\centerline{\includegraphics[width=1.0\columnwidth]{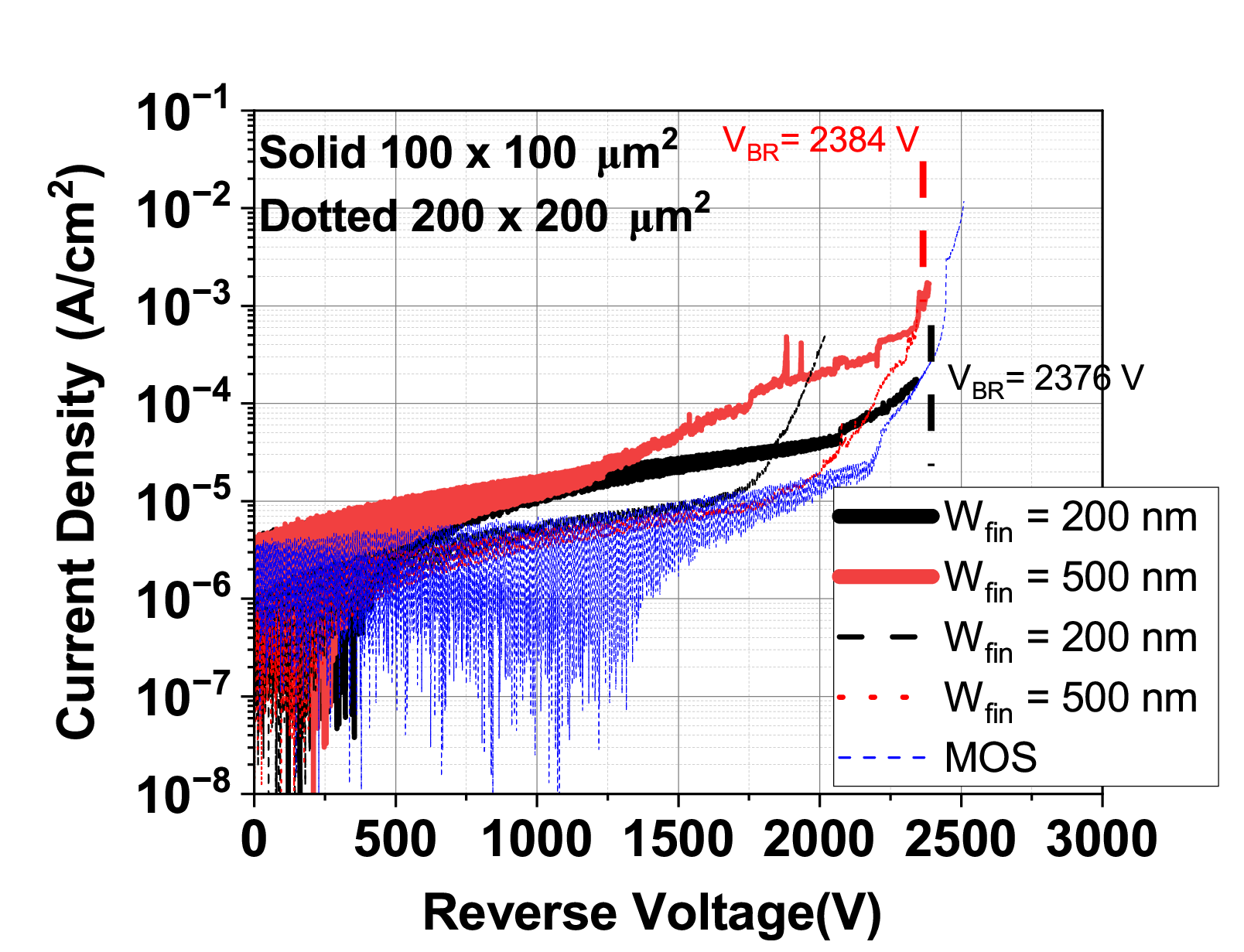}}
\vspace*{-2mm}
\caption{(a) Breakdown measurement for 100 $\mu$m $\times$ 100 $\mu$m and  200 $\mu$m $\times$ 200 $\mu$m  device carried out on the trench diodes with different fin width. Dotted blue in the figure shows the MOS breakdown.}
\label{fig4 BV}
\end{figure}

 \begin{figure*}
\centerline{\includegraphics[width=1.8\columnwidth]{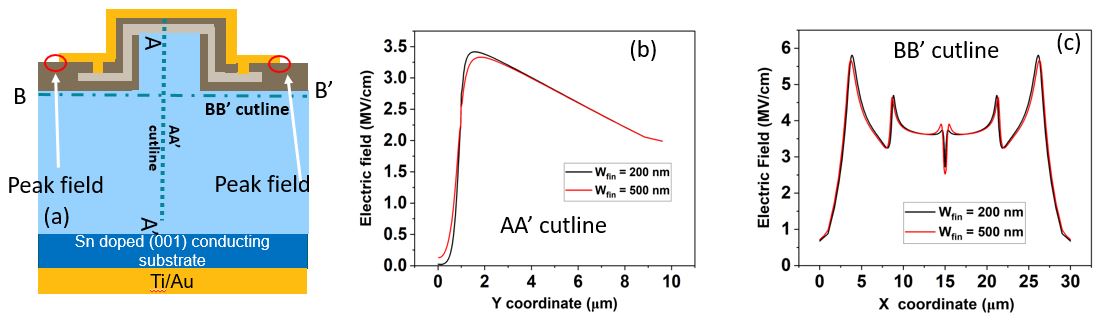}}
\vspace*{-2mm}
\caption{(a) The cross section schematic structure for TCAD Silvaco simulation plot showing AA' and BB' cutline (b) Electric field profile for AA' cutline along vertical direction in the middle of the fin (c) Electric field profile for BB' cutline along the horizontal line at the bottom of the fin in the Ga\textsubscript{2}O\textsubscript{3} layer.}
\label{fig5simulation}
\end{figure*}

High-voltage CV measurement was carried out using B1505 semiconductor parameter analyzer at 1 MHz for extracting the carrier density profile. We have fabricated  Metal Oxide Semiconductor (MOS) test structures with area = 750 $\mu$m $\times$ 750 $\mu$m. Large area pads are essential to measure the capacitance value above the noise floor after reaching punch-through. From CV measurement, we observe that the device capacitance flattens beyond 550 V indicating a punch through of drift layer. The measured punch through voltage also matches closely with simple estimate based on extracted doping profile. We extracted carrier density vs depth profile from C-V measurement and found that the epitaxial drift layer thickness ranges from  8 to 9 $\mu$m before reaching the substrate. The doping profile ranges from 9 $\times$ 10\textsuperscript{15} cm\textsuperscript{-3} to 1 $\times$ 10\textsuperscript{16} from the top of the epitaxial layer to the Sn-doped substrate.

Figure \ref{fig4 BV} shows the reverse current voltage plot of the fabricated device. We submerged the sample in Silicone oil for breakdown measurement. Catastrophic breakdown voltage was measured to be 2.34 kV for 100 $\mu$m $\times$ 100 $\mu$m devices. Device with W\textsubscript{fin} = 200 nm has lower leakage current compared to W\textsubscript{fin} = 500 nm before breakdown indicating enhanced RESURF effect, as expected. While the fin width decreases and the the aspect ratio increases, enhanced RESURF effect causes the surface electric field at the metal-semiconductor interface to reduce. As a result, we get lower leakage for W\textsubscript{fin} = 200 nm device. Larger area device (200 $\mu$m $\times$ 200 $\mu$m) shows similar catastrophic breakdown around 2.34 kV. Both W\textsubscript{fin} = 200 nm and 500 nm devices have similar breakdown and leakage. Generally, larger area devices (200 $\mu$m $\times$ 200 $\mu$m) have a higher probability of encountering defect-related leakage paths dominating over any difference in RESURF effect. We also fabricated some MOS test structures on our sample. MOS test structure breaks down around 2.34 kV with very low leakage before breakdown. This indicates that trench diode breakdown voltage is limited by MOS breakdown. 
We used sub-micron fin width to enhance the RESURF effect, but the final device breakdown is limited by MOS breakdown at the trench bottom. High-k dielectric such as BTO\cite{roy2023ultra,roy2024low}  or thicker high-quality dielectrics, are preferable to enhance the MOS breakdown voltage. Thick field oxides at the trench bottom could further enhance the overall breakdown voltage. \cite{roymulti}

TCAD Silvaco simulation was performed based on the fabricated structure after incorporating field plate. We calculated the dielectric constant of the nanolaminate (Al\textsubscript{2}O\textsubscript{3}/TiO\textsubscript{2}) by CV measurement from accumulation (10 V forward bias). CV measurement on different diameters (60 - 300 $\mu$m) resulted in a dielectric constant of 18 from accumulation capacitance. Based on this dielectric constant, we performed the simulation using 9.5 x 10\textsuperscript{15}cm\textsuperscript{-3} uniform doping 
and 8.5 $\mu$m drift layer at 2400 V reverse bias voltage. We plotted vertical cutline (AA') of the electric field at the middle of the fin in the $\mathrm{\beta} $-{Ga\textsubscript{2}O\textsubscript{3}} drift layer. The electric field at metal-semiconductor junction is very low for W\textsubscript{fin}= 200 nm, and it reaches around 3.4 MV/cm at the drift region. (Fig. \ref{fig5simulation} (b)). Although the electric field at the metal-semiconductor junction is slightly higher for W\textsubscript{win} = 500 nm, overall leakage current is still very low, which is evident from the breakdown plot. (Fig \ref{fig4 BV}). Smaller W\textsubscript{fin} is also beneficial for a smaller
electric-field peak at the trench bottom corners as seen from the Fig. \ref{fig5simulation} (c) along BB' cutline. The parallel plane electric field reaches 3.8 MV/cm (Fig \ref{fig5simulation} (c)) for 2400 V reverse bias. Because of field plate structure, peak field is located at the field plate edge in the dielectric which is shown by red circle in the cross-section schematic. 

We have achieved a power figure of merit (PFoM) of 0.867 GW/cm\textsuperscript{2} for 100 x 100 $\mu$m\textsuperscript{2} device with 2375 V breakdown voltage and 6.5  m$\mathrm{\Omega}$-cm\textsuperscript{2} differential specific on resistance (R\textsubscript{on,sp}) without considering current spreading. PFoM is 0.56 GW/cm\textsuperscript{2} for R\textsubscript{on,sp} = 9.8  m$\mathrm{\Omega}$-cm\textsuperscript{2} after considering current spreading. Both higher breakdown and lower reverse leakage are preferable for high-voltage power amplifiers, and our trench diode performance is compatible with the reported values in the literature.

In summary, we have fabricated a trench Schottky barrier diode with nanoscale fin width using a self-aligned planarization technique and etch back based process. We achieved 140-160 A/cm\textsuperscript{2} on current density with 10\textsuperscript{10} rectification ratio, and low R\textsubscript{on,sp} around 9.8-12 m$\mathrm{\Omega}$-cm\textsuperscript{2} . We measured a catastrophic breakdown at 2300-2400 V for both 100 $\times$ 100 $\mu$m\textsuperscript{2} and  200 $\times$ 200 $\mu$m\textsuperscript{2} devices, which led to a power figure of Merit of 0.56 and 0.85 GW/cm\textsuperscript{2} with and without current spreading, respectively. The leakage current density before breakdown is very low, which can be attributed to the RESURF effect from lower W\textsubscript{fin} and higher aspect ratio of the trench geometry. TCAD simulation shows that the drift region has a peak electric field of 3.4 MV/cm at 2400 V reverse bias voltage.  The final breakdown voltage is limited by MOS breakdown at the trench bottom. We used a dielectric comprising thin layers of TiO\textsubscript{2} and Al\textsubscript{2}O\textsubscript{3} to get a dielectric constant of 17, which shows noise floor leakage upto catastrophic breakdown. Incorporating a high-k dielectric that is compatible with the self-aligned photoresist planarization technique and implementing this process for a low-doped, thicker drift layer is crucial for getting a breakdown voltage exceeding 3 kV.

\begin{acknowledgments}
  This material is based upon work supported by the ARPA-E ULTRAFAST program ( DE-AR0001824 ) and the Microelectronics Commons Program, a DoD initiative, under award number N00164-23-9- G059.   A portion of this work was performed at the UCSB Nanofabrication Facility, an open access laboratory. The MRL Shared Experimental Facilities are supported by the MRSEC Program of the NSF under Award No. DMR 2308708; a member of the NSF-funded Materials Research Facilities Network (www.mrfn.org)

\end{acknowledgments}

\section*{Data Availability Statement}

The data that supports the findings of this study are available from the corresponding author upon reasonable request.

\nocite{*}
\bibliographystyle{ieeetr}
        \bibliography{aipsamp}

\providecommand{\noopsort}[1]{}\providecommand{\singleletter}[1]{#1}%
\begin{thebibliography}{10}

\bibitem{green2022beta}
A.~J. Green, J.~Speck, G.~Xing, P.~Moens, F.~Allerstam, K.~Gumaelius, T.~Neyer, A.~Arias-Purdue, V.~Mehrotra, A.~Kuramata, {\em et~al.}, ``$\beta$-{G}allium oxide power electronics,'' {\em APL Materials}, vol.~10, no.~2, p.~029201, 2022.

\bibitem{ghosh2017ab}
K.~Ghosh and U.~Singisetti, ``Ab initio velocity-field curves in monoclinic {$\mathrm{\beta} $-Ga\textsubscript{2}O\textsubscript{3}},'' {\em Journal of Applied Physics}, vol.~122, no.~3, p.~035702, 2017.

\bibitem{peterson2024200}
C.~Peterson, A.~Bhattacharyya, K.~Chanchaiworawit, R.~Kahler, S.~Roy, Y.~Liu, S.~Rebollo, A.~Kallistova, T.~E. Mates, and S.~Krishnamoorthy, ``200 cm\textsuperscript{2}/vs electron mobility and controlled low 10\textsuperscript{15} cm\textsuperscript{-3} si doping in (010) {$\mathrm{\beta} $-Ga\textsubscript{2}O\textsubscript{3}} epitaxial drift layers,'' {\em Applied Physics Letters}, vol.~125, no.~18, 2024.

\bibitem{blevins2019development}
J.~Blevins, K.~Stevens, A.~Lindsey, G.~Foundos, and L.~Sande, ``Development of large diameter semi-insulating gallium oxide {$\mathrm{\beta} $-Ga\textsubscript{2}O\textsubscript{3}} substrates,'' {\em IEEE Transactions on Semiconductor Manufacturing}, vol.~32, no.~4, pp.~466--472, 2019.

\bibitem{kuramata2016high}
A.~Kuramata, K.~Koshi, S.~Watanabe, Y.~Yamaoka, T.~Masui, and S.~Yamakoshi, ``High-quality {$\mathrm{\beta} $-Ga\textsubscript{2}O\textsubscript{3}} single crystals grown by edge-defined film-fed growth,'' {\em Japanese Journal of Applied Physics}, vol.~55, no.~12, p.~1202A2, 2016.

\bibitem{sasaki2012device}
K.~Sasaki, A.~Kuramata, T.~Masui, E.~G. Villora, K.~Shimamura, and S.~Yamakoshi, ``Device-quality {$\mathrm{\beta} $-Ga\textsubscript{2}O\textsubscript{3}} epitaxial films fabricated by ozone molecular beam epitaxy,'' {\em Applied Physics Express}, vol.~5, no.~3, p.~035502, 2012.

\bibitem{sasaki2014growth}
K.~Sasaki, M.~Higashiwaki, A.~Kuramata, T.~Masui, and S.~Yamakoshi, ``Growth temperature dependences of structural and electrical properties of {Ga\textsubscript{2}O\textsubscript{3}} epitaxial films grown on {$\mathrm{\beta} $-Ga\textsubscript{2}O\textsubscript{3}} substrates by molecular beam epitaxy,'' {\em Journal of Crystal Growth}, vol.~392, pp.~30--33, 2014.

\bibitem{murakami2014homoepitaxial}
H.~Murakami, K.~Nomura, K.~Goto, K.~Sasaki, K.~Kawara, Q.~T. Thieu, R.~Togashi, Y.~Kumagai, M.~Higashiwaki, A.~Kuramata, {\em et~al.}, ``Homoepitaxial growth of {$\mathrm{\beta} $-Ga\textsubscript{2}O\textsubscript{3}} layers by halide vapor phase epitaxy,'' {\em Applied Physics Express}, vol.~8, no.~1, p.~015503, 2014.

\bibitem{kezeng2018}
K.~Zeng, A.~Vaidya, and U.~Singisetti, ``1.85 {kV} breakdown voltage in lateral field-plated {$\mathrm{\beta} $-Ga\textsubscript{2}O\textsubscript{3}},'' {\em IEEE Electron Device Letters}, vol.~39, no.~9, pp.~1385--1388, 2018.

\bibitem{tetzner2019lateral}
K.~Tetzner, E.~B. Treidel, O.~Hilt, A.~Popp, S.~B. Anooz, G.~Wagner, A.~Thies, K.~Ickert, H.~Gargouri, and J.~W{\"u}rfl, ``Lateral 1.8 {kV} {$\mathrm{\beta} $-Ga\textsubscript{2}O\textsubscript{3}} {MOSFET} with 155 {MW}/cm\textsuperscript{2} {P}ower {F}igure of {M}erit,'' {\em IEEE Electron Device Letters}, vol.~40, no.~9, pp.~1503--1506, 2019.

\bibitem{arkka1}
A.~Bhattacharyya, S.~Roy, P.~Ranga, C.~Peterson, and S.~Krishnamoorthy, ``High-{M}obility {T}ri-gate $\mathrm{\beta} $-{Ga\textsubscript{2}O\textsubscript{3}} mesfets with a power figure of merit over 0.9 {GW}/cm\textsuperscript{2},'' {\em IEEE Electron Device Letters}, vol.~43, no.~10, pp.~1637--1640, 2022.

\bibitem{li2019single}
W.~Li, K.~Nomoto, Z.~Hu, T.~Nakamura, D.~Jena, and H.~G. Xing, ``Single and multi-fin normally-off {$\mathrm{\beta} $-Ga\textsubscript{2}O\textsubscript{3}} vertical transistors with a breakdown voltage over 2.6 {kV},'' in {\em 2019 IEEE International Electron Devices Meeting (IEDM)}, pp.~12--4, IEEE, 2019.

\bibitem{farzana2023vertical}
E.~Farzana, S.~Roy, N.~S. Hendricks, S.~Krishnamoorthy, and J.~S. Speck, ``Vertical {P}t{O}\textsubscript{x}/{P}t/$\mathrm{\beta} $-{Ga\textsubscript{2}O\textsubscript{3}} schottky diodes with high permittivity dielectric field plate for low leakage and high breakdown voltage,'' {\em Applied Physics Letters}, vol.~123, no.~19, 2023.

\bibitem{qin20232}
Y.~Qin, M.~Porter, M.~Xiao, Z.~Du, H.~Zhang, Y.~Ma, J.~Spencer, B.~Wang, Q.~Song, K.~Sasaki, {\em et~al.}, ``2 {kV}, 0.7 {m$\Omega$} cm\textsuperscript{2} vertical {$\mathrm{\beta} $-Ga\textsubscript{2}O\textsubscript{3}} superjunction schottky rectifier with dynamic robustness,'' in {\em 2023 International Electron Devices Meeting (IEDM)}, pp.~1--4, IEEE, 2023.

\bibitem{hao2023improved}
W.~Hao, F.~Wu, W.~Li, G.~Xu, X.~Xie, K.~Zhou, W.~Guo, X.~Zhou, Q.~He, X.~Zhao, {\em et~al.}, ``Improved vertical $\mathrm{\beta} $-{Ga\textsubscript{2}O\textsubscript{3}} schottky barrier diodes with conductivity-modulated p-{NiO} junction termination extension,'' {\em IEEE Transactions on Electron Devices}, vol.~70, no.~4, pp.~2129--2134, 2023.

\bibitem{li20237}
J.-S. Li, C.-C. Chiang, X.~Xia, H.-H. Wan, F.~Ren, and S.~Pearton, ``7.5 k{V}, 6.2 {GW} cm\textsuperscript{-2} {NiO}/$\mathrm{\beta} $-{Ga\textsubscript{2}O\textsubscript{3}} vertical rectifiers with on--off ratio greater than 10\textsuperscript{13},'' {\em Journal of Vacuum Science \& Technology A}, vol.~41, no.~3, 2023.

\bibitem{wan20243}
J.~Wan, H.~Wang, C.~Zhang, Y.~Li, C.~Wang, H.~Cheng, J.~Li, N.~Ren, Q.~Guo, and K.~Sheng, ``3.3 k{V}-class {NiO}/$\mathrm{\beta} $-{Ga\textsubscript{2}O\textsubscript{3}} heterojunction diode and its off-state leakage mechanism,'' {\em Applied Physics Letters}, vol.~124, no.~24, 2024.

\bibitem{wang20212}
Y.~Wang, H.~Gong, Y.~Lv, X.~Fu, S.~Dun, T.~Han, H.~Liu, X.~Zhou, S.~Liang, J.~Ye, {\em et~al.}, ``2.41 k{V} vertical {p-NiO}/n-{Ga\textsubscript{2}O\textsubscript{3}} heterojunction diodes with a record baliga's figure-of-merit of 5.18 {GW/cm\textsuperscript{2}},'' {\em IEEE Transactions on Power Electronics}, vol.~37, no.~4, pp.~3743--3746, 2021.

\bibitem{kalarickal2021beta}
N.~K. Kalarickal, Z.~Xia, H.-L. Huang, W.~Moore, Y.~Liu, M.~Brenner, J.~Hwang, and S.~Rajan, ``{$\mathrm{\beta}$-(Al\textsubscript{0.18}Ga\textsubscript{0.82})\textsubscript{2}O\textsubscript{3}} double heterojunction transistor with average field of 5.5 {MV}/cm,'' {\em IEEE Electron Device Letters}, vol.~42, no.~6, pp.~899--902, 2021.

\bibitem{sahabeta2}
C.~N. Saha, A.~Vaidya, A.~F. M. A.~U. Bhuiyan, L.~Meng, S.~Sharma, H.~Zhao, and U.~Singisetti, ``{Scaled $\mathrm{\beta} $-{Ga\textsubscript{2}O\textsubscript{3}} thin channel {MOSFET} with 5.4 MV/cm average breakdown field and near 50 GHz f\textsubscript{MAX}},'' {\em Applied Physics Letters}, vol.~122, no.~18, p.~182106, 2023.

\bibitem{liu202310}
H.~Liu, Y.~Wang, Y.~Lv, S.~Han, T.~Han, S.~Dun, H.~Guo, A.~Bu, and Z.~Feng, ``10-k{V} lateral {$\mathrm{\beta} $-Ga\textsubscript{2}O\textsubscript{3}} {MESFET}s with {B} ion implanted planar isolation,'' {\em IEEE Electron Device Letters}, vol.~44, no.~7, pp.~1048--1051, 2023.

\bibitem{zhou2025c}
M.~Zhou, H.~Zhou, S.~Alghamdi, G.~Gao, X.~Liu, M.~Xiang, X.~Chen, Z.~Liu, S.~Wasly, Y.~Hao, {\em et~al.}, ``C-band $\mathrm{\beta} $-{Ga\textsubscript{2}O\textsubscript{3}} -on-{SiC} rf power mosfets with high output power density and low microwave noise figure,'' {\em IEEE Transactions on Electron Devices}, vol.~72, no.~6, pp.~2874--2878, 2025.

\bibitem{moser2020toward}
N.~Moser, K.~Liddy, A.~Islam, N.~Miller, K.~Leedy, T.~Asel, S.~Mou, A.~Green, and K.~Chabak, ``Toward high voltage radio frequency devices in $\mathrm{\beta} $-{Ga\textsubscript{2}O\textsubscript{3}},'' {\em Applied Physics Letters}, vol.~117, no.~24, 2020.

\bibitem{saha2025high}
C.~N. Saha, N.~J. Nipu, and U.~Singisetti, ``High performance vacuum annealed {$\mathrm{\beta}$-(Al\textsubscript{x}Ga\textsubscript{1-x})\textsubscript{2}O\textsubscript{3}/Ga\textsubscript{2}O\textsubscript{3}} {HFET} with f\textsubscript{T}/f\textsubscript{MAX} of 32/65 {GH}z,'' {\em Applied Physics Express}, vol.~18, no.~7, p.~071001, 2025.

\bibitem{roy2024low}
S.~Roy, B.~Kostroun, Y.~Liu, J.~Cooke, A.~Bhattacharyya, C.~Peterson, B.~Sensale-Rodriguez, and S.~Krishnamoorthy, ``Low {Q\textsubscript{C}V\textsubscript{F}} 20{A}/1.4 {kV} $\mathrm{\beta} $-{Ga\textsubscript{2}O\textsubscript{3}} vertical trench high-k resurf schottky barrier diode with turn-on voltage of 0.5 {V},'' {\em IEEE Electron Device Letters}, vol.~45, no.~12, pp.~2487--2490, 2024.

\bibitem{roy2023ultra}
S.~Roy, B.~Kostroun, J.~Cooke, Y.~Liu, A.~Bhattacharyya, C.~Peterson, B.~Sensale-Rodriguez, and S.~Krishnamoorthy, ``Ultra-low reverse leakage in large area kilo-volt class $\mathrm{\beta} $-{Ga\textsubscript{2}O\textsubscript{3}} trench schottky barrier diode with high-k dielectric resurf,'' {\em Applied Physics Letters}, vol.~123, no.~24, 2023.

\bibitem{li20182}
W.~Li, Z.~Hu, K.~Nomoto, R.~Jinno, Z.~Zhang, T.~Q. Tu, K.~Sasaki, A.~Kuramata, D.~Jena, and H.~G. Xing, ``2.44 kv {Ga\textsubscript{2}O\textsubscript{3}} vertical trench schottky barrier diodes with very low reverse leakage current,'' in {\em 2018 IEEE International Electron Devices Meeting (IEDM)}, pp.~8--5, IEEE, 2018.

\bibitem{li2019field}
W.~Li, K.~Nomoto, Z.~Hu, D.~Jena, and H.~G. Xing, ``Field-plated {Ga\textsubscript{2}O\textsubscript{3}}trench schottky barrier diodes with a {BV\textsuperscript{2}/R\textsubscript{on,sp}} of up to 0.95 {GW}/cm\textsuperscript{2},'' {\em IEEE Electron Device Letters}, vol.~41, no.~1, pp.~107--110, 2019.

\bibitem{sasaki2017first}
K.~Sasaki, D.~Wakimoto, Q.~T. Thieu, Y.~Koishikawa, A.~Kuramata, M.~Higashiwaki, and S.~Yamakoshi, ``First demonstration of $\mathrm{\beta} $-{Ga\textsubscript{2}O\textsubscript{3}} trench mos-type schottky barrier diodes,'' {\em IEEE Electron Device Letters}, vol.~38, no.~6, pp.~783--785, 2017.

\bibitem{dhara2023beta}
S.~Dhara, N.~K. Kalarickal, A.~Dheenan, S.~I. Rahman, C.~Joishi, and S.~Rajan, ``$\mathrm{\beta} $-{Ga\textsubscript{2}O\textsubscript{3}} trench schottky diodes by low-damage ga-atomic beam etching,'' {\em Applied Physics Letters}, vol.~123, no.~2, 2023.

\bibitem{dhara2025charge}
S.~Dhara, A.~Dheenan, and S.~Rajan, ``Charge recovery by vacuum annealing in $\mathrm{\beta} $-{Ga\textsubscript{2}O\textsubscript{3}} multi-fin trench schottky barrier diodes,'' {\em APL Electronic Devices}, vol.~1, no.~2, 2025.

\bibitem{bhattacharyya2022high}
A.~Bhattacharyya, S.~Roy, P.~Ranga, C.~Peterson, and S.~Krishnamoorthy, ``High-mobility tri-gate $\mathrm{\beta} $-{Ga\textsubscript{2}O\textsubscript{3}} mesfets with a power figure of merit over 0.9 {GW}/cm\textsuperscript{2},'' {\em IEEE Electron Device Letters}, vol.~43, no.~10, pp.~1637--1640, 2022.

\bibitem{li2010giant}
W.~Li, O.~Auciello, R.~N. Premnath, and B.~Kabius, ``Giant dielectric constant dominated by maxwell--wagner relaxation in {Al\textsubscript{2}O\textsubscript{3}/TiO\textsubscript{2} }nanolaminates synthesized by atomic layer deposition,'' {\em Applied physics letters}, vol.~96, no.~16, 2010.

\bibitem{lee2013interface}
G.~Lee, B.-K. Lai, C.~Phatak, R.~S. Katiyar, and O.~Auciello, ``Interface-controlled high dielectric constant {Al\textsubscript{2}O\textsubscript{3}/TiO\textsubscript{x}} nanolaminates with low loss and low leakage current density for new generation nanodevices,'' {\em Journal of Applied Physics}, vol.~114, no.~2, 2013.

\bibitem{yates2022demonstration}
L.~Yates, B.~P. Gunning, M.~H. Crawford, J.~Steinfeldt, M.~L. Smith, V.~M. Abate, J.~R. Dickerson, A.~M. Armstrong, A.~Binder, A.~A. Allerman, {\em et~al.}, ``Demonstration of> 6.0-k{V} breakdown voltage in large area vertical {GaN} pn diodes with step-etched junction termination extensions,'' {\em IEEE Transactions on Electron Devices}, vol.~69, no.~4, pp.~1931--1937, 2022.

\bibitem{roy20232}
S.~Roy, A.~Bhattacharyya, C.~Peterson, and S.~Krishnamoorthy, ``2.1 kv (001)-$\mathrm{\beta} $-{Ga\textsubscript{2}O\textsubscript{3}} vertical schottky barrier diode with high-k oxide field plate,'' {\em Applied Physics Letters}, vol.~122, no.~15, 2023.

\bibitem{li2020guiding}
W.~Li, K.~Nomoto, Z.~Hu, D.~Jena, and H.~G. Xing, ``Guiding principles for trench schottky barrier diodes based on ultrawide bandgap semiconductors: a case study in $\mathrm{\beta} $-{Ga\textsubscript{2}O\textsubscript{3}},'' {\em IEEE Transactions on Electron Devices}, vol.~67, no.~10, pp.~3938--3947, 2020.

\bibitem{islam2022defect}
A.~E. Islam, C.~Zhang, K.~DeLello, D.~A. Muller, K.~D. Leedy, S.~Ganguli, N.~A. Moser, R.~Kahler, J.~C. Williams, D.~M. Dryden, {\em et~al.}, ``Defect engineering at the {Al\textsubscript{2}O\textsubscript{3}}/(010) $\mathrm{\beta} $-{Ga\textsubscript{2}O\textsubscript{3}} interface via surface treatments and forming gas post-deposition anneals,'' {\em IEEE Transactions on Electron Devices}, vol.~69, no.~10, pp.~5656--5663, 2022.

\bibitem{roymulti}
S.~Roy, C.~N. Saha, C.~Peterson, W.~J. Mitchell, J.~S. Speck, and S.~Krishnamoorthy, ``Multi-fin $\mathrm{\beta} $-{Ga\textsubscript{2}O\textsubscript{3} Vertical FinFET with Interfin Field Oxide Exhibiting a Breakdown Voltage of 1.8 k{V} and Power Figure of Merit of 1{GW}/cm\textsuperscript{-2}},'' {\em Authorea Preprints}, 2025.

\bibitem{li20181230}
W.~Li, Z.~Hu, K.~Nomoto, Z.~Zhang, J.-Y. Hsu, Q.~T. Thieu, K.~Sasaki, A.~Kuramata, D.~Jena, and H.~G. Xing, ``1230 v $\mathrm{\beta} $-{Ga\textsubscript{2}O\textsubscript{3}} trench schottky barrier diodes with an ultra-low leakage current of< 1 $\mu$a/cm\textsuperscript{2},'' {\em Applied Physics Letters}, vol.~113, no.~20, 2018.

\end{thebibliography}
\end{document}